\documentclass[a4paper,twoside]{article}

\usepackage{epsfig}
\usepackage{subcaption}
\usepackage{calc}
\usepackage{amssymb}
\usepackage{amstext}
\usepackage{amsmath}
\usepackage{amsthm}
\usepackage{multicol}
\usepackage{pslatex}
\usepackage{apalike}
\usepackage{algorithm2e}
\usepackage[bottom]{footmisc}
\usepackage{graphicx}
\usepackage{color}
\usepackage{balance}
\usepackage{SCITEPRESS}     

\begin{document}

\title{Student views in AI Ethics and Social Impact}

\author{\authorname{Tudor-Dan Mihoc\sup{1}\orcidAuthor{0000-0003-2693-1148}, Manuela-Andreea Petrescu\sup{1}\orcidAuthor{0000-0002-9537-1466} and Emilia-Loredana Pop\sup{1}\orcidAuthor{0000-0002-4737-4080}}
\affiliation{\sup{1}Department of Computer Science, Babes Bolyai University, Cluj-Napoca, Romania} 
\email{tudor.mihoc@ubbcluj.ro, manuela.petrescu@ubbcluj.ro, emilia.pop@ubbcluj.ro}
}


\abstract{An investigation, from a gender perspective, of how students view the ethical implications and societal effects of artificial intelligence is conducted, examining concepts that could have a big influence on how artificial intelligence may be taught in the future. For this, we conducted a survey on a cohort of 230 second-year computer science students to reveal their opinions. The results revealed that AI, from the student's perspective, will significantly impact daily life, particularly in areas such as medicine, education, or media. Men are more aware of potential changes in Computer Science, autonomous driving, image and video processing, and chatbot usage, while women mention more the impact on social media.
Both men and women perceive potential threats in the same manner, with men more aware of war, AI-controlled drones, terrain recognition, and information war. 
Women seem to have a stronger tendency towards ethical considerations and helping others.}

\onecolumn \maketitle \normalsize \setcounter{footnote}{0} \vfill

\section{\uppercase{Introduction}}
\label{sec:introduction}

The exposure of AI in the media causes deep unrest in society about the future of this technology. Suddenly, more aware of AI tools \cite{vieweg2021ai}, people begin to debate their implications on the economy, the job market, education, and, not least, ethical issues. Researchers in \cite{ouchchy2020ai} examine and categorize how these topics are portrayed in the press to better understand how this representation could affect public opinion. According to their findings, the media coverage is still on the surface but has a realistic and pragmatic perspective. Other topics related to certain AI methods of information retrieval exposed are: privacy, transparency, biases, censorship, filter bubbles, security, accessibility, data handling, surveillance, job displacement, or data ownership \cite{borenstein2021emerging}.

There is concern that these systems can reinforce or magnify the biases present in the training data, resulting in injustice and discrimination. Furthermore, the processing of user data raises privacy problems because it can lead to data breaches, illegal access, and surveillance \cite{banciu2022ai}.

Students, as future workers and decision makers, should have a high level of understanding of AI ethics, which will help them design and implement AI systems ethically. Therefore, an insight into students' awareness of ethical concerns related to AI is very important.

In the study \cite{Cernadas2020GPersp} the author acknowledges the existence of gender discriminatory biases in the development of students. In order to limit these, he underlines the need to introduce gender perspectives in studies.

Such disparities can also be found in other countries. In some of Romania's largest universities, approximately $31\%$ of computer science students are female. We asked related data to the number of women/men that graduated computer science using a law of transparency that states that public institutions must provide any public information.

There have been studies \cite{pop2024,pop2024p} on the inclination of middle school students to technical education education using machine learning, but little research has been conducted on the students' inclination to actually use and perfect AI.

A survey could reveal the knowledge gaps and capture the diverse perspectives on this topic with respect to gender. The purpose of the study was to contribute to academic research on AI ethics and can provide empirical evidence for scholarly publications, furthering the understanding of ethical considerations in AI.

In order to address these topics, we defined the research questions as:
    \begin{itemize}
        \item [$\star$] Which domains will be most affected by AI? gender-based perspective.
        \item [$\star$] What are the ethical considerations related to the potential threats associated with Artificial Intelligence? 
        \item [$\star$] Who is willing to sacrifice ethical values for money and social status?  Is there a difference between how women and men perceive them?        
    \end{itemize}
We run a survey among students studying computer science in their second year at Babes-Bolyai University, Romania, to get their opinions. We highlight that at the end of the second term, the students have a fundamental understanding of artificial intelligence. They were knowledgeable with rule-based systems, machine learning, decision trees, artificial neural networks, deep learning, intelligent systems, support vector machines, clustering, and problem solving as a search. Throughout the semester, a lecture focused on fraud prevention and AI-related ethical issues. We highlighted sensitive topics and approaches from the perspective of an IT expert, using examples of fake news and the techniques used to create them, as well as how an AI could be tainted and biased through the training database. The survey questions were related to the lecture topics, allowing us to assess students' understanding of the material as well as their thoughts on the ethical implications of AI.

\begin{table*}[t]
    \centering
    \begin{tabular}{|l|l|}
    \hline
        \ \ Q1 \ \ & Please specify your gender. Choice of male, female, or I prefer not to answer. \\
        \hline
        \ \  Q2 \ \ & In which domains do you think Artificial Intelligence will have the greatest impact \\
                & and why? (Mention at least 3) \\
        \hline
        \ \ Q3 \ \ &  Which are considered the main benefits related to 
             Artificial Intelligence that\\  &  might appear in the next 3 years? \\
        \hline
         \ \ Q4 \ \ &  Which are considered the main risks related to 
             Artificial Intelligence that\\  &  might appear in the next 3 years? \\
         \hline
         \ \ Q5 \ \ &  Which are the main reasons you do NOT like a career in Artificial\\  & Intelligence?   \\
          \hline
        \ \ Q6 \ \ &  Which are the main reasons you would like to have a career in \\  & Artificial Intelligence?   \\
         \hline
    \end{tabular}
    \caption{Survey Questions}
    \label{tab:questions}
\end{table*}

\section{Literature review}

Gender disparities related to IT and, more recently, artificial intelligence have been the subject of several studies that have surfaced over the years.

The impact of artificial intelligence on female employment is multifaceted. Men and women are disproportionately affected by industrial growth and platformization, driving the latter out of the labor force \cite{mohla2021material}.
A possible reason for perpetuating stereotypes and discrimination is the diminished importance of women in AI development and implementation.
The need to address the gender gap before it becomes pervasive and embedded in AI culture was highlighted by \cite{Roopaei2021Women}.

According to \cite{abdulllah2019artificial}, AI is seen as a complex potential threat that could affect human behavior, replace jobs, and generate economic inequality. Cultural differences, discrimination, and indiscriminate use of computing resources are just a few ethical challenges that exacerbate these threats \cite{baeza2022ethical}. In \cite{fisher2023fresh} the authors call for increased cooperation and consideration of human interests in the development of artificial intelligence. They pleaded for legislative approaches that address these concerns.

More ethical concerns about AI, such as bias, unemployment, and socioeconomic inequality, are also raised by \cite{green2018ethical}, who also emphasizes the need for a more thorough analysis of these problems.

There are differences in the ethical decision models used by men and women \cite{schminke1997asymmetric}, with a significant difference in the way men and women perceive and prioritize ethical values in different scenarios. When financial gain and social status are involved,  research consistently shows that women are more averse to ethical compromises \cite{kennedy2013willing,kennedy2014willing} than men. They tend to associate business with immorality more than men, and this aversion seems to be related to their lower representation in high-ranking business positions. 

While the underrepresentation of women and minorities in the IT workforce is a problem, their inclusion may offer a way to address the industry's skills gap \cite{gallivan2006workforce}.

Particularly in AI jobs, there are large variations between different groups. \cite{jakesch2022different} discovered that AI practitioners had different values in mind than the broader public, with black and female respondents giving ethical AI ideals more weight. This aligns with the findings of \cite{callan1992predicting}, who indicated that female workers were more inclined to consider discriminatory actions to be morally wrong. According to \cite{rothenberger2019relevance,zhou2020survey}, ethical standards for AI are crucial and may reflect the values that black and female respondents found most important.

\cite{cave2023makes,brown2022attrition,yarger2020algorithmic} also found similar findings, demonstrating that minorities and women are notably underrepresented in AI development jobs. 
One major problem with AI teams is the attrition of individuals with marginalized identities; many people leave because of the culture and environment of these teams. 
This issue is made worse by algorithmic bias in talent acquisition tools, which keeps hiring practices unfair.  

These studies collectively suggest that AI's influence on gender is multifaceted, with the potential to exacerbate existing inequalities.

With this research, we want to have a better understanding of these issues in relation to gender. By addressing the research questions, we want to determine the students' areas of interest in ethical concerns.   
\section{Study Design}

\label{sec:studyDesign}
We used the guidelines provided by \cite{Runeson} to organize this research in accordance with the norms of the scientific community, as stated in \cite{ACM}. We first determined the study's scope before deciding on the methodology.

\textit{Scope}: The purpose of the study was to evaluate students' attitudes about the advancement of artificial intelligence, as well as their interest in learning more about and working in this field with respect to gender identity.

\textit{Who}: Second-year students from computer science departments who took an introductory AI course constituted the group of participants.

\textit{When}: At the conclusion of the semester, we asked the students to take a survey to find out their perspectives.

\textit{How to}: We used a hybrid strategy, analyzing the gathered replies both quantitatively and qualitatively. 

\textit{Observations}: Analysing the answers, we found out that students have concerns about the ethics of AI, so we performed an analysis to find out AI related ethics interests.

\textit{Participation}: Participation in the study was voluntary, and the survey was anonymous, so we could not map participants with their answers.

\subsection {Survey Design}

After we established the purpose of the investigation and formulated the research questions, we prepared the survey questions. The process was an iterative one; first, we elaborated on a set of questions, then two authors discussed, proposed, and validated changes to the form and structure of the questions. The second draft was discussed with the third author, and we agreed on the final versions of the questions. We decided to add questions that we will not use in the study, for example, questions Q3 and Q6. The purpose of these questions was to prevent bias in the received responses and to ask for positive and negative aspects.

We decided to use both closed and open questions. The first question was used to determine the groups and categories of participants (men versus women). The other questions were used to encourage students to freely express their opinions, since open-ended questions can offer valuable insights into the students' perspectives and perceptions. The questions asked in the survey are listed in Table \ref{tab:questions}.

\begin{table*}[ht]
    \centering
     
    \begin{tabular}{|c|c|c|c|c|}
        \hline
       \ \ \textbf{Data Processing} \ \ &\ \ Data Processing\ \  &\ \  Investing \ \  &\ \  Economics \ \  &\ \  Marketing \\ \hline  
                     Men \ \             &\ \ 7.81\%        \ \  &\ \  7.03\% \ \     &\ \  5.47\% \ \     &\ \ 6.25\% \ \     \\ \hline
                      Women \ \          &\ \ 7.69\%\ \          &\ \  8.97\% \ \     &\ \  5.13\% \ \     &\ \ 5.13\% \ \   \\ 
       \hline
    \end{tabular}
   
    \caption{Key item classes percentages for Data Processing}
    \label{tab:tab_class_data_processing}
    
\end{table*}
\subsection{Participants}
The target audience for our survey was formed by second-year computer science students. As a result, the set of participants consisted of 230 individuals, of whom 198 agreed to participate in the study. Seven of them did not wish to disclose their gender, 119 of them were men and 72 women. Given the size of the sample and the fact that the ratio of female participants is similar to the ratio of female students enrolled in the faculty, we may conclude that the study's female representation is statistically significant.

\subsection{Methodology}

The following methodologies used in this study are consistent with those used in previous research efforts \cite{petrescu2023perspective,enase21}. 

At the end of the second semester, we conducted a survey that included open and closed-ended items. Accountable questions facilitate the work with and interpretation of the data, whereas open questions lead to a greater degree of comprehension. The analysis and interpretation of the responses to open questions were carried out using quantitative approaches. The questionnaire surveys met the accepted empirical community norms \cite{ACM}. Thematic analysis was used for text interpretation, following the guidelines provided in \cite{Braun19}. We applied thematic analysis using the following subsequent stages:
\begin{itemize}
    \item [] 
    \begin{enumerate}
        \item Two researchers independently attempted to identify codes or key items within the text.
        \item These key items were divided into classes according to the frequently seen themes or categories. Those with low appearance frequency will be reassigned to larger classes using techniques such as generalization, elimination, and reassignment.
        \item The last step was a comprehensive debate among all authors, so a certain degree of confidence is achieved in the methodology. In the discussions, reviews of several topics, representations, and supporting data for the categorization procedure were included.
    \end{enumerate}

\end{itemize}

We computed the frequency of the important terms. Even if some responses from students were extremely succinct, many others included up to five phrases or justifications. As a result, a response can contain more items or important keys. A direct consequence is that, in our analyses, the total cumulative percentages will exceed 100\%.  

\subsection{Data Collection}

We proposed the questions set in English, as their line of study is English based. The purpose of this choice was to minimize the possible threats resulting from translation. 
The information from the responses was collected exactly as we received it, without any manipulation. 

The students received a link to the anonymous survey via their MSTeams faculty account and were provided with enough time to ensure that their comments were not too brief or non-existent.

\begin{table*}[h]
    \centering
    \begin{tabular}{|c|c|c|c|c|c|c|} 
         \hline
       \ \ \textbf{AM} \ \ &\ \ Engineering\ \  &\ \  Robots \ \  &\ \  Research \ \  &\ \  Physics \ \  &\ \  Architecture \ \  &\ \  Astronomy  \\ \hline  
       Men \ \ &\ \ 3.13\%                           \ \  &\ \  3.91\% \ \  &\ \  1.56\% \ \  &\ \  1.56\% \ \  &\ \  3.91\% \ \  &\ \  2.34\%  \\ \hline
       Women \ \ &\ \ 7.69\%\ \                           &\ \  3.85\% \ \  &\ \  2.56\% \ \  &\ \  2.56\% \ \  &\ \  3.85\% \ \  &\ \  0.00\% \\
       \hline
    \end{tabular}
    \caption{Key item classes percentages for Automatization}
    \label{tab:tab_class_automatization}
\end{table*}

\section{Results}
\label{sec:Results}

In this section, we analyze the data collected and outline the results of student surveys. We took into account the different gender perspectives when we elaborated our analysis. We received 72 responses from students who identified themselves as women, and we considered it a representative number to have valid results. When we calculated the results, we computed the prevalence of key items in women's responses to the total number of women, not the total number of participants in the study.

When we asked students to mention at least three domains (Q1), we had a couple of responses that mentioned four domains, there were three responses that did not mention any domain, all the other students mentioned two or three areas, some of them proving details and explanations. We selected the key items, classified and summarized them, and computed the appearance percentages. Due to this process, the sum of the percentages exceeds 100\%. Responses to the other survey questions were treated in the same way, as each answer could contain zero, one, or more key items, so the total percentages of prevalence of key elements exceed 100\%.

\subsection{\textbf{Q1: Which domains will be most affected by AI? Gender based perspective.}}

To find the answer to this question, we used the answers provided by the students to the survey question: ''In which domains do you think Artificial Intelligence will have the greatest impact and why? (Mention at least 3)''. 

\begin{figure}[!htbp]
    \centering
    \includegraphics[width=1\linewidth]{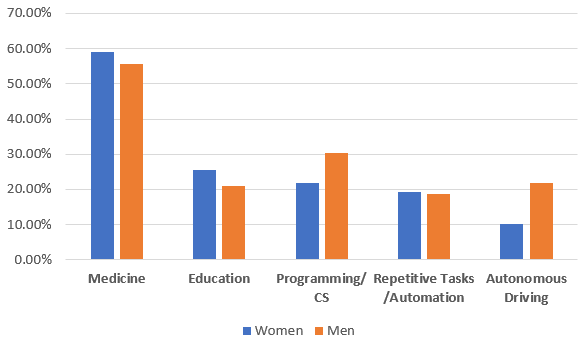}
    \caption{Domains affected by AI. Gender Perspective}
    \label{fig:domains}
\end{figure}

In the perception of the students, the domains with the greatest impact that scored more than 10\% of their responses were medicine, education, programming/computer science (CS), industries, and jobs that involve repetitive tasks and autonomous driving. For the domains most affected, there is no significant difference between the perceptions of men and women, as can be seen in Figure \ref{fig:domains}, only three participants did not answer this question.

\textbf{Medicine} was the domain most mentioned, scoring above 50\% for both men and women: \textit{''health care (detecting diseases, finding the best treatment, etc.)'', ''I think AI will have the greatest impact in medicine, in recognizing flaws unseen by the human eye \& in predictions of many kinds''}. A larger difference appeared for the \textbf{programming/computer science} domain,  considered to be impacted by 30.47\% men and 21.79\% women, \textit{''Programming - helping developers with repetitive code'', ''write code easier'', ''find errors such as missing coma'', ''in testing''}. A larger difference, more than 10\% appears to be related to \textbf{ autonomous driving}, men seem to be more interested in AI's influence, they specify this domain in a percent of 21.88\% compared to women 10.26\%: \textit{''In the automotive industry, with regard to cars that drive themselves'', ''self driving cars''}. \textbf{Education} is mentioned more by women, but the difference is smaller compared to previous domains; only about 4.55\% women mention education more compared to men: \textit{''education as virtual tutoring will be new thing'', ''Learning: you can access all knowledge with AI''}.

The rest of the domain were less prevalent in the student's answers, obtaining less than 10\% and we group them by categories based on the main characteristics: automatization, the capacity to process large amounts of data, and the capacity to generate content.

\textbf{Large data processing capabilities.}
In their opinion, with the ability to process large amounts of data, AI can have an impact in many areas, from analyzing market trends and statistics to making predictions for the stock market and betting: \textit{''Identifying market trends and strategies (economy)'', ''Data analysis, works easily with large data'', ''Stock market prediction, obviously'', ''Betting (sports, betting)''}. Also, concerns about marketing appear: \textit{''because it is easier to detect patterns''} in people's behavior. A particular response synthesized this aspect: \textit{''I think it will have the greatest impact on education, data processing, and research because it has access to information beyond a single person's capacity''}. As shown in Table \ref{tab:tab_class_data_processing}, both genders fairly mentioned these aspects.

\textbf{Automatization.}(AM)
An answer stated that \textit{''Automatization of everyday tasks, can be done better by AI than by humans''}, either if we are referring to \textit{''Surgery domain - great robots for operations''} or to \textit{''repetitive and boring tasks''}. There are no significant gender differences in perceptions, as can be seen in Table \ref{tab:tab_class_automatization}.

\textbf{Content Generation.} 
A specific characteristic of some AI tools is their ability to generate new content in terms of art, music, photo, and video editing and processing.  In art \textit{''because it can generate high quality free photos''} or \textit{''generate an image based on description''}, in \textit{''design by generating images, patterns or logos''}, in fact \textit{''In the creative industry (images, videos, music, etc.), the linguistic/news industry (articles, translations, etc.)''}. Music and even fashion were mentioned: \textit{''music: you can create music with AI'', ''fashion: you can use different AI to create clothes''}. Men mention chatbots more compared to women: \textit{'' Internet bots -  easy to write text that looks written by a human''}, women mentioned more media and social media: \textit{''in algorithms used by social media platforms and in the video game industry because it could learn about human behavior''}.

\begin{table*}[h]
    \centering
    \begin{tabular}{|c|c|c|c|c|c|}
        \hline
       \ \ \textbf{Creation} \ \ &\ \ Art \ \  &\ \  Text\ \     &\ \  Video/photo \ \  &\ \  (Social)  \ \  &\ \ Chatbots \\ 
       \ \     \ \               &\ \      \ \  &\ \  generation \ \  &\ \ editing \ \  &\ \   Media  \ \  &\ \  \\ \hline 
             \ \   Men \ \    &\ \ 6.25\% \ \  &\ \  4.69\% \ \     &\ \  8.59\% \ \     &\ \ 4.69\% \ \  &\ \  10.94\%    \\ \hline
              \ \ Women \ \   &\ \ 5.13\%\ \  &\ \  1.28\% \ \     &\ \  3.85\% \ \     &\ \ 7.69\%\ \  &\ \     2.56\% \\ \hline
    \end{tabular}
    \caption{Key item classes percentages for Content Generation}
    \label{tab:tab_class_content_generation}
\end{table*}

\textbf{Ethics related concerns.} Some responses revealed concerns related to mass manipulation through the detection and use of patterns in people's behavior. In the view of the students, AI poses threats through advanced processing and generation of multimedia content (video, image, or audio) -- \textit{''deep fakes, it is a major problem if they become hard to detect''}. The use of artificial intelligence in communication and social media can have an impact on governments, wars, and politics: \textit{''politics because the generation of different fake videos leads to disinformation''}. Male participants also mentioned the use of drones, war recognition, and fighting.
Although the question had a different scope, there were responses concerned about possible job replacements since automation could eliminate repetitive jobs: \textit{''IT, in testing; I think it will be the first job that will disappear'', ''industrial implies that robots could replace people''}. 

\textbf{Q1 Conclusion.} 
The domains mentioned in the responses were diverse, suggesting that AI will have a non-negligible impact on daily life. It was interesting to analyze the fact that the domains most present in the answers were the domains where there was a lot of progress in AI: medicine, education, or photo/video processing and generating.

In a general view, there are no significant differences between men's and women's perceptions related to the most affected domains; however, men seem more aware of the potential changes in Computer Science, autonomous driving, image and video processing, and chatbot usage (either to replace jobs or to be used in personalized learning). Women mention, compared to men, only the impact on social media. 

\subsection{\textbf{Q2: What are the ethical considerations related to the potential threats associated with Artificial Intelligence?}}

\begin{figure}[!htbp]
    \centering
    \includegraphics[width=1\linewidth]{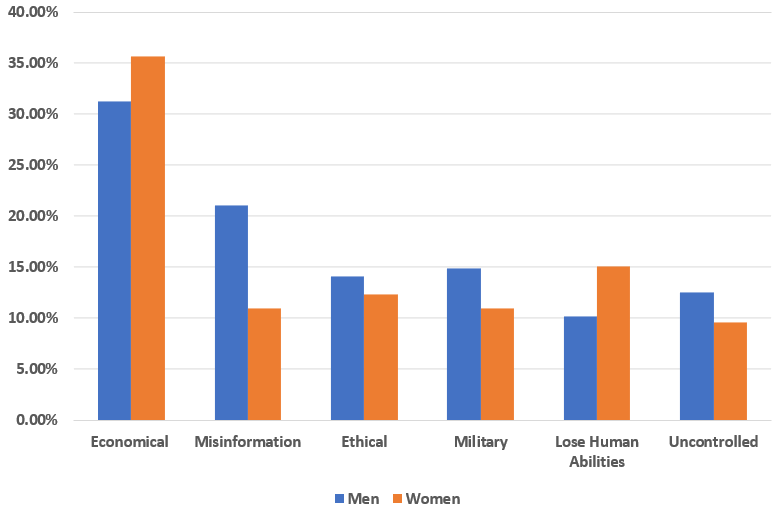}
    \caption{Major threats due to AI. Gender Perspective}
    \label{fig:threats}
\end{figure}

Following the analysis of the responses received, we classified the key items into classes, resulting in five main classes of threats: economic, information-related, general, apocalyptic fear, and absence of risks. Additionally, there were students who did not respond to this question or indicated that they were indifferent to it. A set of responses, 5.24\% of them, emphasized the need for regulations in the field of artificial intelligence. A rather significant percentage of 23.03\% of students did not answer to this question and stated that they consider AI not to pose any risks 2.45\% and other 1.47\% mentioned that they don't care. We designated this question as optional to avoid introducing bias, as we did not want to force students to respond if they had not considered the implications of AI. The main threats identified by us were related to economic issues, misinformation and fake news, ethical concerns, the military, the loss of human abilities, and becoming uncontrolled. The gender-based concerns are presented in Figure \ref{fig:threats}.

\textbf{Economic threats} are represented by a major category: job cuts due to AI 26.56\% men and 34.25\% women, the possible replacement of the artist's work, which was mentioned separately in 4.69\% men's responses and 1.37\% women's responses. Job cuts are exemplified in responses such as: \textit{''loss of jobs due to human replacement by robots'', ''Job loss maybe in programming''} and even in creative domains such as art: \textit{''replacing artists''} work or \textit{''destruction of multiple career paths (arts, writing currently endangered)''}. We got some responses that take into consideration even \textit{''market manipulation''}, but in terms of prevalence, responses that mention other domains are rare.

\textbf{Misinformation}
Misinformation is the second most mentioned threat, with a total of 10.16\% men and 4.11\% women. Separately, the threat of fake news was mentioned by 10.94\% of men and 6.85\% of women. AI can facilitate \textit{''fake news spreading''}, also \textit{''misinformation will be more easily achieved''}, and \textit{''deep fake that becomes more and more convincing''}.

\textbf{Ethical concerns} related to AI are mentioned as ethical implications by 5.48\% women and 8.59\% men, concerns related to copyright and/or plagiarism, and a generic \textit{''bad influence} or \textit{''bad direction''} mentioned by 6.85\% women and 5.47\% men.

Ethical implications are mentioned in a general manner, often quite succinct: \textit{''the problem of ethic'', ''ethics concerns''}, or \textit{''I think there will be legal and moral implications''}, others offer more details: there is \textit{''one enormous risk is that AI will be needed for unethical or malicious purposes''}. Plagiarism is mentioned as \textit{''copyright of train data, defamation, ethical concerns'', ''Art plagiarism, de-commissioning artists, intellectual property theft (intentional or not)''}.

\textbf{Military} The use of AI for military purposes appears in more than 14\% of the answers, mentioned by men and women in relatively similar percentages: 14.84\% men versus 10.96\% women: \textit{''Use in military, become smarter than humans, become incontrollable'', ''Military: can lead to mass destruction''}, can be used for :\textit{''political biases''} and \textit{''propaganda. You can already find on the internet deep fake of presidents''}. Protecting information in the information war is essential, and AI can make a change at individual levels: \textit{''privacy issues, surveillance'', ''ethical concerns regarding personal information and integrity''} or at mass levels: \textit{,''It could get too much access to important features''}, it could \textit{''leak dangerous information''}, and provide a \textit{''better control and influences over masses''}. 

There is a threat that AI could become \textbf{uncontrolled}, 12.50\% men and 9.59\% women expressed this concern; they mention \textit{''Self-improvement of algorithms, to the point where control is lost''}, and that AI could \textit{''become self-conscious''}.

The last concern mentioned by students is not related to ethics but more to \textbf{loss of human abilities} 10.16\% men and 15.07\% women referred to it: \textit{'Decrease in the level of intelligence of humanity'','' We can become dependent on it and lose our ability to think''}.

\textbf{Conclusion}. Men and women perceive potential threats relatively in the same manner; even if the men are more aware of the destructive character, they mention war, AI controlled drones, terrain recognition, or informational war in terms of sensitive information leakage. In addition, men appear to be more aware of the phenomenon of fake news and disinformation. 

\subsection{\textbf{Q3: Who is willing to sacrifice ethical values for money and social status? Is there a difference between how women and men perceive them? }}
When people are asked about themselves, they often respond how they would like to be or act \cite{Rogers}, so instead of asking if they would give up financial benefits and status for ethical reasons, we asked them to provide reasons for which they would follow or not a career path in AI. If the number of students who did not answer the previous questions was relatively small, in these cases a larger number of students had chosen not to answer or had clearly specified that they did not know. When asked to give reasons for not working in AI, 21.92\% women and 34.11\% men did not give any reason, and 17.81\% women and 6.98\% stated that they want another career. When asked for reasons to work in AI, 28.77\% women and 35.16\% men did not provide reasons, since the other 15.07\% women and 9.38\% men clearly specified that they do not want a career in AI.

We grouped the items in the responses into two large categories: ethical and non-ethical reasons. The non-ethical reasons can be classified into financial and non-financial reasons, summarizing more than 45\% for both men and women for both questions. In Table \ref{tab:tab_personal_reasons} we can see the key items for these non-financial personal reasons for choosing or not choosing to work in an AI project for men versus women.

\begin{table*}[h]
    \centering
    \begin{tabular}{|l|l|}
        \hline
        \textbf{Personal reasons for not choosing} & \textbf{Men | Women} \\ 
        \hline
        Complex/Difficult                         & 13.95\% | 12.33\% \\ 
        Math                                      & 13.95\% | 10.96\% \\
        Boring                                    & 5.43\% | 1.33\% \\
        Constantly evolving                       & 0.00\% | 4.11\% \\ 
        No interest                               & 6.20\% | 8.22\% \\ 
        Other career                              & 6.98\% | 17.81\% \\ 
        \hline
        \textbf{Personal reasons for choosing}    & \textbf{Men | Women} \\
        \hline
        Interesting                               & 30.47\% | 31.51\% \\ 
        Great impact                              & 5.47\% | 6.85\% \\ 
        AI is the future                          & 9.38\% | 9.59\% \\ 
        \hline
    \end{tabular}
    \vspace{2pt}
    \caption{Key item classes percentages for personal non-financial reasons of men versus women.}
    \label{tab:tab_personal_reasons}
\end{table*}

The non-ethical reasons are mostly financial - the number of jobs and the opportunities to work in AI, the pay for such a job. Men are more interested in these aspects 18.75\% compared to 12.33\% women: \textit{''Current lack of AI-focused jobs that allow transitioning between workplaces for non-seniors.'', ''Hard to get it'',''it has a very right chance that it will not pay well because you gotta be on top''}.

Ethical reasons are present in 13.70\% of the women's responses and 10.08\% in the men's responses. Some perceive that AI is heading in a bad direction 6.85\% women versus 3.88\% men, and some believe that there are moral problems 4.11\% women versus 3.88\% men. The students who clearly specify that they do not want to contribute to AI's development due to ethical reasons or because the fact job cuts and the effect on the people are 2.72\% women and 2.33\% men.

\textbf{Results Discussion}

There are no significant differences between genders in students' perceptions related to beliefs and perceptions, although there are some minor differences between men and women regarding the mention of more frequent war fair, inaccurate information, financial aspects, and ethical reasons.

Both men and women mentioned that the domains most affected by AI development are the domains that are already using AI (medicine, art, military, automation). A notable difference between men and women appears only in fake news propagation and disinformation, aspects most mentioned by men. Women mention loss of human abilities reasons more than men. When asked about the reasons for a career in AI, women more mentioned the desire to help: 9.59\% vs 6.25\%: \textit{''I could save people from a rare disease'', ''To help people with this technology in each aspect of life to make a lot of comfort in society and help the planet be more healthy''}.
Another difference is the fact that men are more preoccupied by the financial aspects (number of job opportunities and money) and they are less preoccupied about ethics, compared to women, who are more preoccupied about ethics, they want more "to help".

\section{Threats to Validity}

Through the analysis and application of the community standards outlined in \cite{ACM}, our objective was to minimize any possible risks.
Similar practices have also been applied in other research papers, as in \cite{PetrescuPopMihoc2023,Kiger}. 
We also mitigated the potential threats to validity that were identified in software engineering research \cite{ACM}.
Due to the guidelines considered, we have identified and analyzed three aspects: construct validity, internal validity, and external validity. For internal validity, we focus specifically on the participant set, participant selection, dropout contingency measures, and author biases.

\textbf{Construct validity.} To reduce the authors' biases, the questions were developed in a multi-step process as outlined in the Survey Design. The suggested survey questions aligned with the research objectives as stated in the Introduction. 

\textbf{Internal validity.} 
The potential internal threats identified by us were participant and participant selection, drop-out rates, author subjectivity, and ethics.

\begin{itemize}
    \item []  \textbf{Participant set and participant selection.} Every student enrolled in the AI course, regardless of gender or other characteristics, has been notified about the survey and asked to participate in it. Consequently, the target group of participants was comprehensive, eliminating any potential risks associated with the set of participants or their selection.

    \item [] \textbf{Drop-outs rates.}
    Due to the voluntary nature of the survey, we had limited methods to reduce the dropout rates. 
    The survey consisted of only a few questions in order to increase student participation. The open-ended questions were optional.
    By outlining the benefits of our research and allowing the survey to remain open for two weeks, we encouraged participation.  

    \item [] \textbf{Author subjectivity.} 
    Aware of the possible subjectivity in data processing, we have taken this into account and examined it.
    We tried to reduce this risk by using text analysis according to the recommended data processing protocols.  
    Additionally, by taking into account suggested data processing practices, we have validated each other's work.
    By debating every facet (including the clarity of the approach, the selection of keywords, and the themes), we attempted a non-subjective approach.
    
    \item [] \textbf{Ethics in our research}. 
    We demonstrated our commitment to ethics by providing participants with information about our purpose of collecting data, our anonymous data collection method, and our intended use of the data. Additionally, we made it clear that participation was voluntary, and some of them chose not to do so as evidence (from 230 students, only 198 participated).

\end{itemize}

\textbf{External validity.}
We examine the potential to generalize the findings of our study. One concern is whether the results can be extrapolated to a broader cohort of AI (or even IT) students. We mention that any extrapolation can not be made to the whole society since we considered a specific cohort. However, we can extrapolate to the students' set enrolled in IT, with some caution, because the percentage of women who participated in the study is comparable to the percentage of women enrolled in general in Computer Science studies in universities. Also, the men/women percentages in our study correlate with the global gender percentages in STEM according to \textit{the Global Gender Gap Report} (2023) \cite{GlobalGenderGapReport2023}.

\section{Conclusion and Future Work}
\label{sec:Conclusion}

This research aims to evaluate students' attitudes and interest in AI advancement and ethics with respect to gender identity. The survey design was iterative, with changes made to ensure accuracy. The data were collected in English to minimize translation threats and ensure accurate analysis.

The study involved 230 second-year computer science students, of which 198 participated. The study's female representation was statistically significant, as it correlates with the gender percentages of women enrolled in computer science in universities. The methodology used was thematic analysis, with key items divided into classes based on frequency.

The responses to the survey revealed that AI will significantly impact daily life, particularly in areas such as medicine, education, and photo/video processing. While there are no significant differences between men and women in their perceptions of the most affected domains, men are more aware of potential changes in Computer Science, autonomous driving, image and video processing, and chatbot usage. Women, however, mention more the impact of losing human abilities.

Both men and women perceive potential threats in the same manner, with men more aware of war, AI controlled drones, terrain recognition, and informational war. They also seem to be more aware of fake news and disinformation.

Ethical reasons were more prominent among women, with women expressing a desire to help people in various aspects of life, such as saving people from rare diseases and making society more comfortable. Men were more preoccupied with financial aspects, while women were more concerned with helping others.

The study aimed to minimize potential risks and validate the findings through construct validity, internal validity, and external validity.

The study revealed that both men and women had distinct motivations and priorities when it came to emerging technologies. While men focused more on financial gains and advancements, women had a stronger inclination towards ethical considerations and helping others. Overall, the research shed light on the gender differences in motivations and highlighted the need for a balanced approach in the development and implementation of emerging technologies.

In the future, we hope to expand the study to include a larger and more diverse cohort of students from other universities, resulting in a widely comparable data set across European or international institutions. Analyzing the data not only in terms of gender but also comparing differences among different nationalities as well as the development over time for several generations of students would be great next steps, and the paper at hand is the first step in that direction.

\balance

\bibliographystyle{apalike}

\bibliography{example,bib_GAP,ethics_g_01}

\end{document}